\documentclass[aps,prl,twocolumn,groupedaddress,floatfix]{revtex4}

\usepackage{amsmath}
\usepackage{amsfonts}
\usepackage{amssymb}
\usepackage{graphicx}

\begin{document}

\title{Comment on ``Scaling of the anomalous Hall effect in Sr$_{1-x}$Ca$_{x}$RuO$_{3}$''}


\author{Yevgeny Kats}
\altaffiliation{Present address: Department of Physics, Harvard University, Cambridge, MA 02138, USA.}
\author{Lior Klein}
\affiliation{Department of Physics, Bar-Ilan University, Ramat-Gan 52900,
Israel}

\maketitle

Mathieu \textit{et al.} \cite{Mathieu} measured the anomalous Hall effect (AHE)
$\rho_{xy}$ and the magnetization $M$ of Sr$_{1-x}$Ca$_{x}$RuO$_{3}$ films as a
function of temperature, and claimed that the transverse conductivity
$\sigma_{xy}$ of samples with different doping levels scales with the
temperature- and doping-dependent magnetization $M(T,x)$, thus providing strong
support to their recently suggested interpretation of the AHE in SrRuO$_{3}$ as
an intrinsic effect due to Berry phase monopoles in $\mathbf{k}$-space
\cite{Fang}. In this Comment we point out that the Letter does not provide such
support, since the presented data are problematic, and that even if the data
are trusted, the results allow for scaling of comparable quality assuming
$\rho_{xy}$ scales with $\rho$, as might be expected for conventional
(extrinsic) models for AHE.

When SrRuO$_{3}$ films are grown on $(001)$ substrates of SrTiO$_{3}$, they may
grow in six different orientations relative to the substrate. For two of the
orientations, the easy axis of the magnetization is in the film plane, and for
the other four orientations -- roughly at $45^{\circ}$ out of the film plane
\cite{Marshall}. Mathieu \textit{et al.} report that their SrRuO$_{3}$ films
have perpendicular magnetic anisotropy (and do not provide any evidence for its
intrinsic nature \cite{perpendicular}). Since perpendicular anisotropy does not
correspond to any of the intrinsic orientations, the films are likely to
contain grains with more than one orientation, which would introduce
uncontrolled errors in measuring the intrinsic magnetization and AHE.

\begin{figure}[ptb]
\includegraphics[scale=0.32, trim=160 320 180 -150]{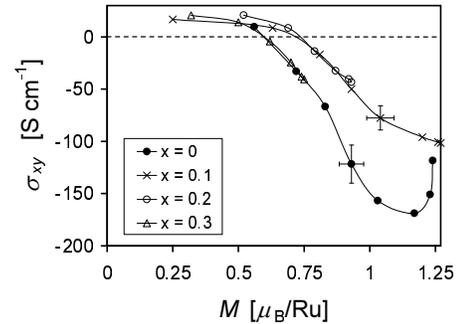}
\caption{Scaling of the transverse conductivity $\sigma_{xy}$. (Reproduced from
Fig. 3 of Ref. \cite{Mathieu}. Fewer data are shown.)}%
\label{sigma-scaling}
\end{figure}

It should be noticed that the curves of $\sigma_{xy}(M)$ for the different
samples (reproduced in Fig. \ref{sigma-scaling}) are separated by much more
than the indicated error bars. Such differences cannot be attributed to
uncertainty in film thicknesses (which is one of the error sources mentioned by
the authors), since while for $M = 1$ $\mu_{B}/$Ru, $\sigma_{xy}$ of $x=0$ is
by $100\%$ larger than of $x=0.1$, at $M = 1.25$ $\mu_{B}/$Ru it is only by
$20\%$ larger (thickness uncertainty would scale the results by a constant
factor). The differences cannot be attributed to doping-driven changes in
material parameters, since while the curves of $x=0.1$ and $0.2$ are very
different from the curve of $x=0$, the curve of $x=0.3$ almost coincides with
the curve of $x=0$. On the other hand, uncontrolled differences in the amount
of twinning between samples may account for such messy behavior as a function
of doping.

\begin{figure}[ptb]
\includegraphics[scale=0.32, trim=160 330 180 -130]{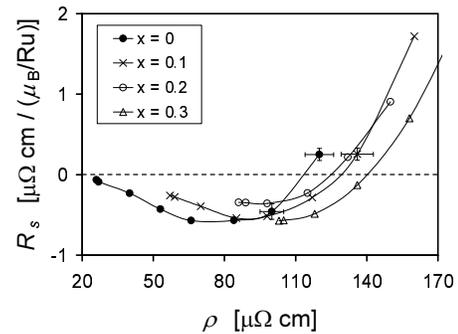}
\caption{Data extracted from Ref. \cite{Mathieu} and presented in terms of the
AHE coefficient $R_{s}$ as a function of the resistivity $\rho$.}%
\label{Rs-scaling}
\end{figure}

Furthermore, even if the data are trusted, the large spread of the data does
not allow to distinguish between the suggested (intrinsic) interpretation and
the conventional (extrinsic) models for AHE. For example, Fig. \ref{Rs-scaling}
shows that if the data of Ref. \cite{Mathieu} are re-analyzed in terms of a
na\"{\i}ve conventional model, i.e. $\rho_{xy}=R_{s}(\rho)M$, a scaling of
comparable quality is achieved.

We would like to note that there are additional reasons to doubt the
applicability of the Berry phase model to the AHE in SrRuO$_{3}$, and they are
presented in Ref. \cite{EHE-decrease}.

We acknowledge support by the Israel Science Foundation founded by the Israel
Academy of Sciences and Humanities.

\end{document}